\documentclass[pra,twocolumn,amsmath]{revtex4}
\usepackage{amssymb}

\usepackage{graphicx}
\usepackage{dcolumn}
\usepackage{bm}

\begin{document}

\title{Spin-Exchange-Relaxation-Free Magnetometry\\
Using Elliptically-Polarized Light}
\author{V. Shah}
 \email{vishal.k.shah@gmail.com}
\author{M. V. Romalis}%
\affiliation{%
Department of Physics, \\Princeton University, Princeton,\\ New Jersey 08544, USA}%

\begin{abstract}
Spin-exchange relaxation free alkali-metal magnetometers typically operate
in the regime of high optical density, presenting challenges for simple and
efficient optical pumping and detection. We describe a high-sensitivity Rb
magnetometer using a single elliptically-polarized off-resonant laser beam.
Circular component of the light creates relatively uniform spin polarization
while the linear component is used to measure optical rotation generated by
the atoms. Modulation of the atomic spin direction with an oscillating
magnetic field shifts the detected signal to high frequencies. Using a
fiber-coupled DFB laser we achieve magnetic field sensitivity of 7 fT/$\sqrt{%
\mathrm{Hz}}$ with a miniature $5\times5\times5$ mm Rb vapor cell.
\end{abstract}

\maketitle


\affiliation{Department of Physics, \newline
Princeton University, Princeton,\newline
New Jersey 08544, USA}



\section{Introduction}
Alkali-metal magnetometers have recently achieved magnetic field
sensitivity on the order of 1 fT/$\sqrt{\mathrm{Hz}}$, previously
possible only with superconducting quantum interference devices
(SQUID) operating in liquid helium \cite{BudkerRomalis}. High
sensitivity is obtained by operating at a high alkali-metal density
of $10^{13}-10^{15}$ cm$^{-3}$ in a spin-exchange relaxation free
(SERF) regime. In this regime the rate of alkali-metal spin-exchange
collisions is much greater than the Zeeman precession frequency,
resulting in the reduction of the spin-exchange relaxation of the
Zeeman coherences \cite{Happer}. Magnetic field sensitivity in the
femto-tesla range opens a number of new applications previously
accessible only for SQUID magnetometers
\cite{Fagaly06,ledbetter2008}. Of particular interest is detection
of biological magnetic fields from the heart or the brain \cite
{Matti93,Gratta}, presently the biggest area of SQUID use. In this
application tens to hundreds of SQUID sensors are placed around the
subject to create a spatial map of the magnetic field. Detection and
mapping of brain magnetic fields has been demonstrated with a SERF
magnetometer using a large vapor cell and a multi-channel
photodetector \cite{Xia}. However the flexibility of magnetic field
mapping would be greatly improved if one could instead use a
collection of simple, compact and self-contained atomic
magnetometers while still maintaining high magnetic field
sensitivity.

Here we describe such compact, fiber-coupled atomic magnetometer.
Our approach combines several novel features necessary for achieving
high magnetic field sensitivity in a small detection volume. It
operates in the regime of high optical depth of several hundred on
resonance, which presents challenges for efficient optical pumping
and detection. In the conventional configuration \cite {Kominis02},
a circularly polarized pump beam is tuned to the center of the D1
resonance to polarize the atoms and a linearly-polarized probe beam
is tuned off-resonance to measure their optical rotation. The pump
beam can propagate through the optically-thick vapor only by pumping
most of the atoms into the dark stretched state. However, this
reduces the sensitivity of the magnetometer, since it is maximized
for a spin polarization of 50\%. By detuning the pump laser
off-resonance so that the optical depth is less than one it is
possible to maintain the polarization of the alkali-metal vapor
closer to the optimum value. For a pressure-broadened optical
resonance the efficiency of optical pumping is not reduced when the
laser is detuned away from the resonance.

Detection of spin polarization is still performed using polarization
rotation of the light, which is generally the most efficient
``non-demolition'' technique of spin interrogation. It is also
largely immune from laser intensity noise. The simplest method of
light polarimetry involves a polarizing beam-splitter cube at
45$^{\circ}$ to the initial polarization and two balanced
photodetectors. However, low-frequency sensitivity of this technique
is usually degraded due to motion of the laser beam. Better
performance at low frequency can be achieved if the polarization of
the light is modulated, using, for example, a separate Faraday
modulator \cite{Kominis02}. However, this would not be practical in
a compact, self-contained magnetometer. Therefore we use
alkali-metal atoms themselves to modulate the polarization of the
light. The direction of the atomic spin polarization is modulated by
a large angle using an oscillating magnetic field, which in turn
creates a large modulation of the light polarization. The signal
proportional to the DC magnetic field is measured at the modulation
frequency. We discuss the requirements for the choice of the
modulation frequency so the vapor remains in the SERF regime. For
sufficiently high alkali-metal density, the modulation frequency of
several kHz can be used without significant loss in sensitivity,
largely eliminating low-frequency noise associated with beam motion.

To reduce the overall complexity of the magnetometer we use a single
elliptically polarized laser beam for both pumping and probing. This
reduces the efficiency of optical pumping by $\sqrt{2}$ and also
reduces the optical rotation signal by $\sqrt{2}$. The light from a
DFB laser is send to the sensor through a multi-mode optical fiber.
Photodetectors are incorporated into the magnetometer. Off-resonant
partially circularly polarized light causes a significant light
shift and we discuss methods to mitigate its effects. We demonstrate
magnetic field sensitivity of 7 fT/$\sqrt{\mathrm{Hz}}$, limited by
Johnson noise in the magnetic shields. The measurements are
performed in a vapor cell with dimensions of $5\times5\times5$ mm.
This roughly represents an optimal magnetometer size for
non-invasive measurements on human subjects since magnetic field
sources of interest are typically submerged several centimeters
below the skin.

Earlier work on simplifying the operation of SERF magnetometers using spin
modulation has been presented in \cite{walker06,vshah07}. In \cite{walker06}
a magnetic field modulation technique was used to shift the measured signal
to higher frequencies. However, the relationship between the modulation
frequency and the requirement for operation in the SERF regime was not
discussed and separate pump and probe beams were used. In \cite{vshah07} a
single circularly polarized laser beam tuned to resonance was used for both
pumping and probing the vapor. The optical density of the alkali-metal vapor
was restricted to be on the order of unity to avoid excessive absorption of
the laser beam. The noise in the detection of the optical transmission was
dominated by laser intensity noise. The magnetic field sensitivity
demonstrated here is about an order of magnitude greater than in either of
these earlier papers.

In Section II we discuss the theory of atomic magnetometery using
elliptically polarized light and magnetic field modulation in the SERF
regime. In Section III we describe the experimental implementation of the
magnetometer, present comparison with theory and measurements of the
magnetic field sensitivity.

\section{Theory}

\begin{figure}[tbp]
\includegraphics[width=8 cm]{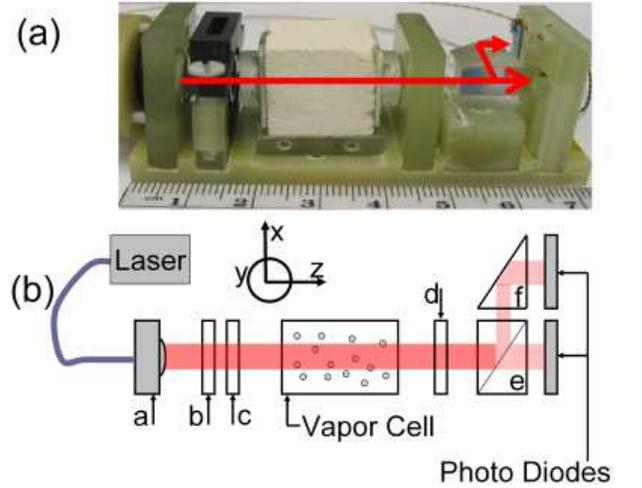}
\caption{(a) Assembled magnetometer sensor (side view). (b) A
schematic of magnetometer arrangement (top view). \textbf{a} is a
fiber holder and a collimator, \textbf{b} is a linear polarizer,
\textbf{c} is a quarter waveplate, \textbf{d} is a focusing lens and
a half wave plate combination, \textbf{e} is polarizing beam
splitter and \textbf{f} is a prism. The angle between the polarizer
axis and the fast axis of the quarter wave is equal to $\pi$/8 and
the angle between the polarizer axis and the half wave plate axis is
equal to $\pi$/4.} \label{f1}
\end{figure}

We consider propagation of the light in an optical system shown in Figure
\ref{f1}, representing the experimental arrangement used in this work. The
light from a multi-mode fiber passes through a plate linear polarizer and a
quarter-waveplate with its optic axis oriented at an angle $\theta $
relative to the linear polarizer. The propagation of the light through the
Rb atomic vapor is conveniently analyzed in the circular polarization basis:
\begin{equation}
\mathbf{E}=E_{0}\left[ Ae^{ik_{r}z}\hat{R}+Be^{ik_{l}z}\hat{L}\right]
e^{-2\pi i\nu t},
\end{equation}
where $\hat{R}=(\hat{x}-i\hat{y})/\sqrt{2}$ and $\hat{L}=(\hat{x}+i\hat{y})/%
\sqrt{2}$ are the right and left circular basis vectors and $\nu $ is the
laser frequency. After the combination of linear polarizer and quarter-wave
plate, $A=\left( \cos \theta +\sin \theta \right) /\sqrt{2}$ and $B=\left(
\cos \theta -\sin \theta \right) /\sqrt{2}$ . In the atomic vapor the two
circular components of the light experience different indices of refraction $%
k_{r}=2\pi \nu n_{r}/c$ and $k_{l}=2\pi \nu n_{l}/c,$ which have both real
and imaginary components to account for light retardation and absorption, $%
n_{r}=n_{r}^{r}+in_{r}^{i}$ and $n_{l}=n_{l}^{r}+in_{l}^{i}$. Polarization
of the light after the vapor cell is analyzed using a polarizing
beamsplitter oriented at an angle $\alpha $ relative to the optic axis of
the waveplate and measured with a pair of photodetectors. The sum $\mathcal{S%
}$ and difference $\mathcal{D}$ of the photodetector signals is given by:
\begin{eqnarray}
\mathcal{D} &=&E_{0}^{2}e^{-d(n_{l}^{i}+n_{r}^{i})}\cos [2\alpha
+d(n_{l}^{r}-n_{r}^{r})]\cos 2\theta \\
\mathcal{S} &=&\frac{E_{0}^{2}}{2}\left[ \left( 1+\sin 2\theta \right)
e^{-2dn_{r}^{i}}+\left( 1-\sin 2\theta \right) e^{-2dn_{l}^{i}}\right] ,
\end{eqnarray}
where $d=2\pi \nu l/c$ and $l$ is the length of the cell.

We consider here optical pumping on the D1 line in an alkali metal vapor in
the presence of high buffer gas pressure. Therefore, the absorption profile
is given by a pressure-broadened Lorentzian and the interaction (both
absorption and dispersion) between atoms and photons is proportional to $(1-%
\mathbf{s}\cdot \mathbf{P})$ where $\mathbf{s}$ is the photon spin $\mathbf{s%
}=i\mathbf{E}\times \mathbf{E}^{*}/E_{0}^{2}$ and $\mathbf{P}=\langle
\mathbf{S}\rangle /S$ is the electron spin polarization \cite{Appelt99,Happer72}%
. Hence the indices of refraction are given by
\begin{eqnarray}
n_{r} &=&1+\kappa (1+P_{z})L(\nu )/\nu \\
n_{l} &=&1+\kappa (1-P_{z})L(\nu )/\nu
\end{eqnarray}
where $L(\nu )=1/(\nu _{0}-\nu -i\Delta \nu /2)$ is a complex
Lorentzian profile with a FWHM $\Delta \nu $ centered at $\nu _{0}$.
Here $\kappa =nr_{e}c^{2}f/4\pi$, where $n$ is the alkali-metal
density, $r_{e}$ is the classical electron radius and $f\simeq 1/3$
is the oscillator strength for the D1 transition. For $\alpha =\pi
/4$ we get for the photodiode difference signal $\mathcal{D}$

\begin{equation}
\mathcal{D}=E_{0}^{2}e^{-\sigma nl }\sin \phi \cos 2\theta ,
\label{diff}
\end{equation}
where the absorption cross-section $\sigma = c
r_{e}f\mathrm{Im}[L(\nu )]$ and the optical rotation angle $\phi =c
r_{e}fnlP_{z}\mathrm{Re}[L(\nu )].$ The laser frequency $\nu $ is
detuned off-resonance to generate a finite optical rotation angle
$\phi$  proportional to the spin polarization $P_{z}$. With large
detuning  we also have $\sigma nl\lesssim 1$ to avoid significant
light absorption in an optically-dense alkali-metal vapor. Note that
in the presence of spin polarization $P_{z}$ the light will become
more circularly polarized after propagating through the vapor, but
this affects only the sum of the intensities $\mathcal{S}$, not
their difference $\mathcal{D}$.

\begin{figure}[tbp]
\includegraphics[width=8.5 cm]{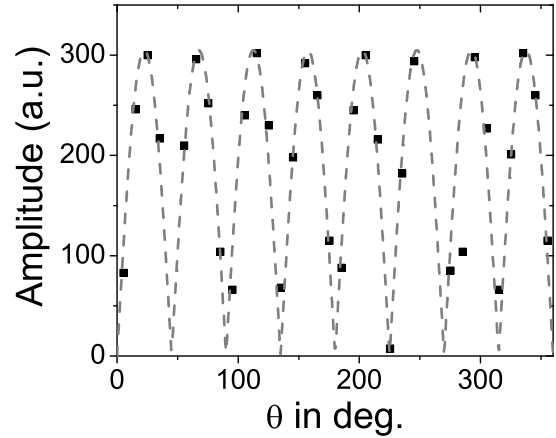}
\caption{(Dots) Experimental data showing amplitude of the zero
field resonance as a function of the angle between the polarizer and
the quarter-waveplate $\theta $. The measurements were made using
low light intensities such that the rotation angle $\phi <<\pi /4$.
(Dash) Numerical fit to the data using a functional form of $|sin(4
\theta)|$ expected for the resonance amplitude in the
low-polarization limit.}\label{f4_1}
\end{figure}

To calculate the spin polarization of the alkali-metal atoms we use Bloch
equations that are valid in the SERF regime where the density matrix assumes
a spin temperature distribution \cite{Allred02,ledbetter08}
\begin{equation}
\frac{d\mathbf{P}}{dt}=D\nabla ^{2}\mathbf{P}+\frac{1}{Q(P)}\left( \gamma
\mathbf{P}\times \mathbf{B}+R(\mathbf{s}-\mathbf{P})-\frac{\mathbf{P}}{T_{2}}%
\right) ,  \label{dp}
\end{equation}
where $D$ is the diffusion constant, $\mathbf{B}$ is the external
magnetic field, $\gamma =2\pi g_{s}\mu _{B}/\hbar =2\pi \times $
$2.8\times 10^{10}$ Hz/T is the electron gyromagnetic ratio and
$Q(P)$ is the nuclear slowing down factor that depends on the spin
polarization \cite{ledbetter08}. $Q(P)=6$ for $^{87}$Rb in the low
polarization limit. $T_{2}$ is the transverse spin relaxation time
in the absence of the light, dominated by spin-destruction collisons
with alkali-metal and buffer gas atoms. One can also approximately
take into account the effects of spin relaxation on cell walls by
additing the decay rate of the fundamental diffusion mode. $R$ is
the optical pumping rate for an unpolarized atom, $R=\sigma \Phi $,
where $\Phi =cE_{0}^{2}/(8\pi h\nu )$ is the photon flux. The photon
spin $\mathbf{s}=-\sin 2\theta \,\hat{z},$ ignoring changes in the
polarization of the light due to propagation in the atomic vapor.
First consider a steady-state solution to Eq. (\ref{dp}) for the
case when $B_{y}$
and $B_{z}$ are equal to zero. We get for $P_{z}$%
\begin{equation}
P_{z}=\frac{sR\tau }{\gamma ^{2}B_{x}^{2}\tau ^{2}+1},  \label{Pz}
\end{equation}
which is a zero-field Lorenztian resonance as a function of $B_{x}$
with a half width at half maximum (HWHM) $\Delta B_{x}=1/\gamma
\tau$, where $\tau =(R+1/T_{2})^{-1}$ is the atomic spin coherence
time. If the optical rotation angle $\phi $ is small, then the
height of the resonance measured in the photodiode difference signal
$\mathcal{D}$ changes with the waveplate angle $\theta $ as $|\sin
4\theta |$. The experimentally measured resonance amplitude in the
limit of low pumping rate and small spin polarization is shown in
Fig. \ref{f4_1} as a function of $\theta $, confirming this
behavior. The waveplate angle $\theta =\pi /8$ gives an optimal
combination of optical pumping and probing with a single beam,
reducing each by a factor
of $\sqrt{2}$. It is also possible to achieve conditions when $\phi >\pi /2$%
, which results in a more complicated signal shape. The central
feature of that shape near $B_{x}=0$ becomes narrower, increasing
magnetometer sensitivity.

\begin{figure}[tbp]
\includegraphics[width=6.5 cm]{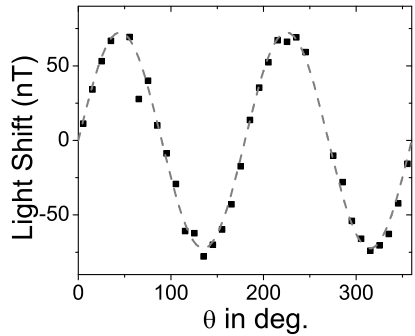}
\caption{ (Dots) Experimentally observed light shift as a function $\theta $
under typical operating conditions. The light-shift was measured by
monitoring the strength of magnetic field in the z-direction required to
minimize the zero-field resonance line-width. (Dash) A fit to the
experimental data using the functional form of the photon spin $%
s=\sin(2\theta )$. The absolute value of the light-shift is
consistent with Eq. (\ref{ls}) within a factor of two, due to
uncertainty in the incident light intensity.} \label{f4_2}
\end{figure}

One negative effect of optical pumping using off-resonant
elliptically-polarized light is that it generates significant
light-shift \cite{Happer67}. The light-shift can be viewed as a
fictitious magnetic field along the direction of light propagation
whose magnitude is given by \cite{Appelt99},
\begin{equation}
\mathbf{B}_{LS}=scr_{e}f\Phi \mathrm{Re}[L(\nu )]/\gamma
\,\hat{z}\label{ls}
\end{equation}
It can be seen by finding a steady-state solution to Eq. (\ref{dp})
with a non-zero $B_{z}$ that the light shift will produce a
broadening of the magnetic field resonance as a function of $B_{x}$.
The light shift can be
compensated by applying an equal and opposite real magnetic field in the $%
\hat{z}$ direction. However, in practice spatial inhomogeneity of
light intensity due to transverse laser beam profile and absorbtion
in the cell produce gradients of $B_{LS}$ that are difficult to
compensate externally and result in residual broadening of the
magnetic resonance. To reduce the effects of light-shift
inhomogeneity we expanded the Gaussian laser beam to a size larger
than the cell and used only the relatively uniform central part of
it. In Figure \ref{f4_2} we show the average value of the
light-shift as a function $\theta ,$ measured by finding a value of
external $B_{z}$ field that gives the narrowest resonance. The
behavior of the light-shift as a function of $\theta $ is consistent
the expected $\sin (2\theta )$ dependance of the photon spin. It is
possible to eliminate the effects of the light shift by periodically
switching the frequency of the laser light to either side of the
resonance on a time scale faster than the atomic spin coherence
time. This would also introduce a high-frequency modulation into the
measured rotation signal.

To use the zero field magnetic resonance for sensitive measurement of $B_{x}$
field we need to generate a dispersion signal with a steep slope centered
around $B_{x}=0$ \cite{LaloE70,Slocum73}. We use a magnetic field modulation
$B_{x}^{tot}=B_{x}+B_{\mathrm{mod}}\cos \omega t$ which also shifts detected
signal to a higher frequency $\omega $ without the need for an external
Faraday modulator or similar techniques. To determine the optimal modulation
parameters we consider the case when $\omega \gg \left( R+1/T_{2}\right) $,
going beyond steady state solution of Eq. (\ref{dp}). The solution of the
Bloch equations in this regime has been obtained in \cite{LaloE70}. In
particular, the component of $P_{z}$ at the modulation frequency $\omega $
is given by
\begin{equation}
P_{z}(\omega )=\frac{2sR\gamma B_{x}\tau ^{2}\sin \omega t}{\gamma
^{2}B_{x}^{2}\tau ^{2}+1}J_{0}\left( \frac{\gamma B_{\mathrm{mod}}}{%
Q(P)\omega }\right) J_{1}\left( \frac{\gamma B_{\mathrm{mod}}}{Q(P)\omega }%
\right)  \label{mod}
\end{equation}
where $J_{0}$ and $J_{1}$ are Bessel functions of the first kind. This gives
a dispersion signal as a function of $B_{x}$ with a maximum slope obtained
when the modulation index $m=\gamma B_{\mathrm{mod}}/Q(P)\omega =1.08$ and $%
R=1/T_{2}=1/2\tau $.

It is generally advantageous to increase the modulation frequency
while simulatanouesly increasing the modulation amplitude
$B_{\mathrm{mod}}$ to move detected signal frequency further from
technical noise sources at lower frequencies. However when
$B_{\mathrm{mod}}$ is sufficiently large, relaxation due to
spin-exchange collisions begins to increase. In the regime when the
spin-exchange rate $R_{se}\gg \gamma B_{x}^{tot}$ and the spin
polarization is relatively low, the additional transverse spin
relaxation due to spin exchange collisions is given by
\cite{savukov05}
\begin{equation}
\frac{1}{T_{2}^{se}(B_{x})}=\frac{5}{18}\frac{\gamma ^{2}(B_{x}^{tot})^{2}}{%
R_{se}}
\end{equation}
for an alkali metal with $I=3/2,$ such as $^{87}$Rb. Since the frequency of $%
B_{x}$ modulation is much larger than the relaxation rate, we can calculate
the average relaxation rate over one modulation cycle
\begin{equation}
\left\langle \frac{1}{T_{2}^{se}}\right\rangle =\frac{\omega }{2\pi }%
\int_{0}^{2\pi /\omega }\frac{1}{T_{2}^{se}(B_{x})}dt=\frac{5}{36}\frac{%
\gamma ^{2}B_{\mathrm{mod}}^{2}}{R_{se}}
\end{equation}
assuming that $B_{x}$ is much smaller than $B_{\mathrm{mod}}.$ In Eq. (\ref
{mod}) the spin coherence time $\tau $ then becomes $\tau =$ $%
(R+1/T_{2}+\left\langle 1/T_{2}^{se}\right\rangle )^{-1}.$ The
condition for maximum modulation frequency that can be used without
loss of sensitivity is set by $\left\langle
1/T_{2}^{se}\right\rangle \lesssim R+1/T_{2}$. This gives a
modulation frequency $\omega \sim \sqrt{R_{se}/5T_{2}},$ which is
much larger than $1/T_{2}$ in the SERF regime. For a given
modulation frequency one can find the optimal modulation amplitude from Eq. (%
\ref{mod}) taking into account changes in $\tau $ due to spin-exchange
relaxation.

We find that for our experimental conditions  the optical rotation angle $%
\phi \propto P_z $ is often larger than $\pi /2.$ In that case the
photodetector difference signal $\mathcal{D}\propto \sin \phi$ is
not a monotonic function of $P_{z}$. The modulation index $m$ can be
adjusted to maximize the first harmonic signal in $\mathcal{D}$. The
slope of the dispersion signal as a function of $B_{z}$ is maximized
at a smaller value of $m$ and continues to increase with atomic path
length $nl$.  We find that in our experimental conditions the
optimal value of the modulation index $m$ is 0.5 to 0.8.

\section{Experimental Setup and Results}

A schematic of the experimental setup is shown in Figure
\ref{f1}(b). The light from a 795 nm DFB laser tuned close to
$^{87}$Rb D1 transition was coupled into a multimode optical fiber
and was delivered to the magnetometer setup. At the entrance of the
magnetometer package, a plano-convex lens was used to collimate the
light after which a sheet linear polarizer was used to define the
polarization axis of the incoming light. Elliptically-polarized
light was created using either a $\lambda /4$ waveplate oriented at
$\theta =\pi /8$ or a $\lambda /8$ waveplate oriented at $\theta
=\pi /4$ relative to the linear polarizer. The beam passed through a
cubical atomic vapor cell containing enriched $^{87}$Rb and
approximately 300 Torr Helium and 100 Torr N$_{2}$ buffer gases. The
inner volume of the vapor cell was 0.09 cm$^{3}$, it was housed in a
boron-nitride oven that was electrically heated using a high
resistance titanium wire. The operating temperature of the cell was
approximately 200$^{\circ }$C. The boron-nitride oven was enclosed
in a low thermal conductivity ceramic to reduce heat loss due to
convection. To prevent magnetic fields generated by the electrical
heaters from interfering with magnetic field measurements, the
heater current was chopped on and off with a duty cycle of 50\% at a
frequency of 0.02-0.2 Hz. The measurements were made while the
heating current was off. In other experiments we have successfully
used AC currents at 20-100 KHz to heat the cell continuously without
affecting magnetic field measurements \cite{Kornack}. The light
transmitted through the vapor cell passed through a half-wave plate
and was analyzed using a polarizing beam splitter and a pair of
silicon photodiodes. The half-wave plate was rotated to balance the
intensity in the two detectors in the absence of optical rotation
from atoms.

\begin{figure}[tbp]
\includegraphics[width=6.5 cm]{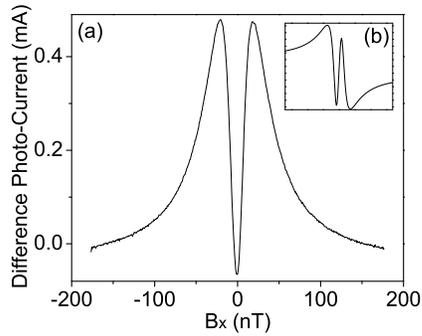}
\caption{ (a) A typical zero field resonance seen as a function of
transverse magnetic field B$_{x}$ when B$_{y}$ and effective B$_{z}$
magnetic fields are zeroed. The resonance has multiple peaks due to
optical rotation angles greater than $\pi /4$. We estimate roughly
40 percent contribution to the line-width comes from residual
light-shifts. (b) Lock-in amplifier dispersion signal. A 2 kHz
sinusoidal magnetic modulation of 200 nT$_{peak}$ amplitude was
applied to $B_x$ to generate the signal.} \label{f2}
\end{figure}
Figure \ref{f2} shows a narrow zero-field SERF resonance signal
obtained by monitoring photodiode difference signal as a function of
B$_{x}$ field. Note that the rotation signal ``turns over" because
it exceeds $\pi /4$. B$_{y}$ field and the sum of B$_{z}$ and
B$_{LS}$ were zeroed by minimizing the linewidth of the resonance.
The average light intensity at the entrance of
the cell was about 40 mW/cm$^{2}$ and the Rubidium density was 8$\times $10$%
^{14}$/cm$^{3}$. The laser was detuned from the resonance by about 45 GHz
where the optical density $\sigma nl$ was close to unity while the optical depth on resonance was 200. The relaxation rate $%
1/T_{2}$ due to a sum of Rb-Rb spin destruction, Rb-buffer gas
spin-destruction and wall collisions was $1/T_{2}=1200$ sec$^{-1}$,
and the optical pumping rate $R=880$ sec$^{-1}.$ The average
light-shift expressed in terms of a fictitious magnetic field
B$_{LS}$ was about 35 nT.
\begin{figure}[tbp]
\includegraphics[width=7 cm]{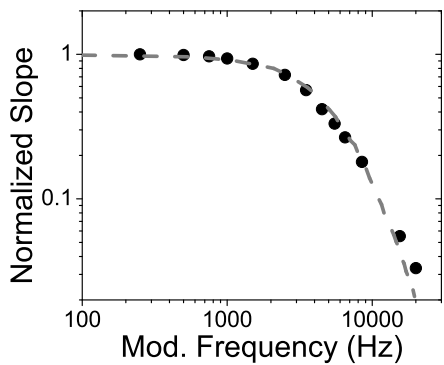}
\caption{(Dots) Normalized lock-in amplifier signal slope around
$B_x=0$ as a function of the modulation amplitude for a fixed ratio
of modulation amplitude to frequency equal to 0.17 nT/Hz which
corresponds to modulation index $m=0.8$ (Dash) Predictions for the
slope based on Eq. (\ref{mod}).} \label{f5}
\end{figure}

To operate as a magnetometer, the zero-field resonance was converted
to a steep dispersive line shape centered around $B_{x}=0$ using
additional modulation of $B_{x}$ field. The output of a lock-in
amplifier with a phase of 90 deg. relative to the modulation is
shown in the inset of Fig. \ref {f2}. The shape of the signal is
more complicated than a simple dispersion curve due to large optical
rotations. The modulation index is adjusted to maximize the slope of
the dispersion curve near $B_{x}=0$. In Fig. \ref{f5} we show the
slope at the zero crossing as a function of the frequency and
amplitude of the modulation field with the modulation index
$m=\gamma B_{\mathrm{mod}}/Q(P)\omega =0.8$ held fixed. The slope
does not change substantially for modulation frequencies up to 2
kHz, about 10 times larger than the bandwidth of the magnetometer,
in agreement with predictions.
\par
\begin{figure}[tbp]
\includegraphics[width=8.5 cm]{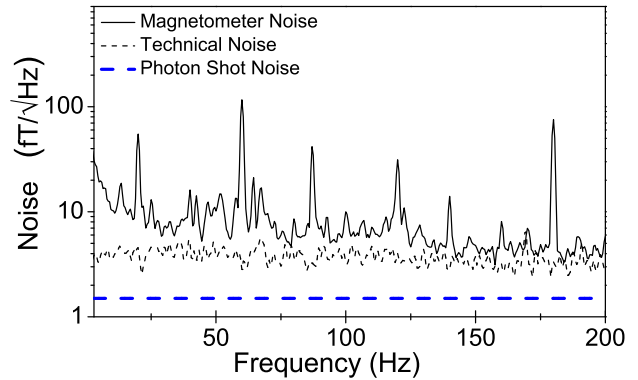}
\caption{(a) (Solid-line) Spectral density of the magnetic field
noise. The 3-dB bandwidth of the magnetometer was 150 Hz.
(Dashed-line) Polarimetry technical noise observed by switching of
the modulation field used for lockin detection. (Thick dashed-line)
Calculated photon shot noise-limited sensitivity of the
magnetometer.} \label{f3}
\end{figure}

In Figure \ref{f3} we show the noise spectrum of the magnetometer
when operated in the neighborhood of B$_{x}=0$ with a modulation
frequency of 2 kHz. Except for spikes at discrete frequencies the
spectrum of the magnetic field noise is limited by Johnson noise in
magnetic shields \cite{sklee}. Polarimetry noise in the absence of
atoms can be tested by turning off magnetic field modulation. It is
slightly lower but still above photon shot noise. Increasing the
modulation frequency to 5 kHz reduces polarimetry noise but also
broadens magnetic resonance due to turn-on of the spin-exchange
relaxation. Operating at even higher Rb density would avoid this
problem.

\section{Conclusion}

In conclusion, we have theoretically and experimentally examined a
simple scheme for implementing SERF magnetometer using a single
off-resonant elliptically polarized light beam. This scheme combines
both simplicity and high performance to achieve femtotesla level
sensitivity using a miniature vapor cell with an inner volume of
0.09 cm$^{3}$. The complete magnetometer package consisting of
optics, heated vapor cell and a pair of photodetectors occupied 40
cm$^{3}$ that we believe can be further reduced by over an order of
magnitude in a relatively straight forward way. A compact sealed
package would also likely improve low-frequency noise of the
magnetometer.  The laser light used here was coupled into the
magnetometer using a multi-mode optical fiber which demonstrates the
ability to operate an array of independent SERF sensors using a
single high power laser. Magnetic field modulation can either be
applied by a common large coil to all sensors \cite{Xia} or
individually to each cell using small coils designed to produce a
zero dipole and quadrupole magnetic moments at large distances. The
sensitivity and size of the magnetometer makes it ideally suited for
magnetoencephalography \cite {Matti93} and other multi-channel
applications that can operate near zero total magnetic field.
\par
This work was supported by an ONR MURI award.



\end{document}